\documentclass[jkps,preprint,fleqn,showpacs,showkeys]{revtex4}

\usepackage{graphicx}
\usepackage{amssymb}
\usepackage{amsmath}
\usepackage{bm}
\begin{document}
\setcounter{page}{0}
\title[]{Electron and hole transmission through superconductor - normal metal interfaces}
\author{Kurt \surname{Gloos}}
\email{kgloos@utu.fi}
\thanks{Fax: +382-2333-5470}

\affiliation{Wihuri Physical Laboratory, Department of Physics   and Astronomy, University of Turku, FIN-20014 Turku, Finland}
\affiliation{Turku University Centre for Materials and Surfaces (MatSurf), FIN-20014 Turku, Finland}

\author{Elina \surname{Tuuli}}
\affiliation{Wihuri Physical Laboratory, Department of Physics and Astronomy, University of Turku, FIN-20014 Turku, Finland}
\affiliation{The National Doctoral Programme in Nanoscience (NGS-NANO), FIN-40014 University of Jyv\"askyl\"a, Finland}

\date{\today}

\begin{abstract}
We have investigated the transmission of electrons and holes through interfaces between superconducting aluminum ($T_c = 1.2\,$K) and various normal non-magnetic metals (copper, gold, palladium, platinum, and silver) using Andreev-reflection spectroscopy at $T = 0.1\,$K. 
We analysed the point contacts with the modified BTK theory that includes Dynes' lifetime as a fitting parameter $\Gamma$ in addition to superconducting energy gap $2\Delta$ and normal reflection described by $Z$. 
For contact areas from $1\,$nm$^2$ to $10000\,$nm$^2$ the BTK $Z$ parameter was 0.5, corresponding to transmission coefficients of about $80\%$, independent of the normal metal. 
The very small variation of $Z$ indicates that the interfaces have a negligible dielectric tunneling barrier. Fermi surface mismatch does not account for the observed transmission coefficient. 
\end{abstract}

\pacs{85.30.Hi, 73.40.-c, 74.45.+c}


\keywords{point contacts, metal interfaces, normal reflection, Andreev reflection}

\maketitle

\section{Introduction}

An interface between two conductors presents an obstacle for charge (electron or hole) transport, transmitting a fraction $\tau$ of the incident current and reflecting the remainder $1-\tau$.
Normal reflection plays a central role in Andreev-reflection spectroscopy according to the BTK theory \cite{Blonder1982} which assumes that the transmitted electrons and the retro-reflected holes are affected in the same way.
This allows to measure the transmission coefficient of normal-superconductor interfaces.
Blonder and Tinkham \cite{Blonder1983} explained the usually observed Andreev reflection double-minimum structure - an enhanced resistance around zero bias inside the energy gap - as being due to a combination of tunneling through a dielectric layer and the mismatch of Fermi velocities. 
A dielectric oxide \cite{Simmons1964} or water/ice layer \cite{Repphun1995} has to be expected when the contact between the two conductors is not prepared at ultrahigh-vacuum conditions.
Even a contact between two identical metals disrupts the crystal lattice symmetry and should lead to some amount of normal reflection.
Describing the real interface with a $\delta$-function barrier and in the one-dimensional free electron approximation \cite{Landau1977}, the transmission coefficient $\tau = 1 / (1+Z^2)$ can be obtained from \cite{Blonder1983}
\begin{equation}
  Z^2 = Z_b^2 + (1-r)^2/(4 r)
  \label{Z parameter}
  \end{equation}
where $r = v_{F1}/v_{F2}$ is the ratio of Fermi velocities $v_{F1}$ and $v_{F2}$ of the two electrodes.
Thus one could directly measure Fermi-velocity ratios once the contribution $Z_b$ of the dielectric barrier is known.

This seemingly simple situation changed dramatically with the discovery of heavy-fermion superconductors where the 'heavy' conduction electrons form the Cooper pairs \cite{Steglich1979}.
The first point-contact study of those compounds by U.~Poppe \cite{Poppe1985} and Steglich {\it et al.} \cite{Steglich1985} focussed on Giaever-type tunneling to measure the density of states of the new superconductors and the Josephson effect to probe the symmetry of the heavy-fermion order parameter, without considering Andreev reflection.
E.~W.~Fenton \cite{Fenton1985} predicted a huge normal reflection coefficient of interfaces between a heavy-fermion and a conventional metal, corresponding to $Z \gg 1$, because of their very small Fermi velocities down to $1/1000$ of that of conventional metals.
This idea got partial support by a large background residual resistance of those heavy-fermion contacts where the cross-sectional area could be determined independently \cite{Gloos1995,Gloos1996a,Gloos1996b}. 
However, the corresponding tunneling-like Andreev reflection anomalies have never been found.

A second approach by Deutscher and Nozi{\`e}res \cite{Deutscher1994} seemed to resolve the question of the puzzlingly small normal reflection. 
According to them not the heavy particles cross the interface, but the bare ones.
This suggests that it is not the mismatch of Fermi velocities but that of the Fermi wave numbers that matters for normal reflection.
Equation \ref{Z parameter} remains with $r$ the ratio of Fermi wave numbers.
For interfaces between heavy-fermion compounds and conventional metals this ratio is of order unity, and therefore the $Z$ parameter should be rather small.
Since then many point-contact Andreev reflection experiments on heavy-fermion superconductors have been performed, for example \cite{DeWilde1994,Park2008}, focussing on the symmetry of the superconducting order parameter, and paying less attention to normal reflection.

Meanwhile, the study of the proximity effect at superconducting - normal metal thin film layered structures, which depends strongly on the transparency of the interfaces, has progressed \cite{Attanasio2006,Kushnir2009}.
Such thin films are usually fabricated in ultra-high vacuum, making a dielectric interface barrier unlikely and leaving Fermi surface mismatch to explain normal reflection.
These experiments suggest that $\tau \lesssim 0.5$ (corresponding to $Z \gtrsim 1$) for contacts between simple metals, considerably less than the expected $\tau \approx 1$ ($Z \approx 0$) in free-electron approximation.
One can also measure directly the current perpendicular to plane (CPP) resistance of an interface with a well defined geometry \cite{Pratt2009,Sharma2009} and compare it with electronic-structure calculations \cite{Xu2006,Xu2006b}. 
The CPP resistance should contain information about normal reflection, but for us it is difficult to extract.

Measuring spin polarization using Andreev-reflection spectroscopy \cite{Soulen1998} is another research topic that relies heavily on normal reflection.
According to the general point of view \cite{Bugoslavsky2005,Baltz2009}, the true spin polarization is only obtained at highly transparent interfaces when $Z \rightarrow 0$ while the measured polarization drops to zero around $Z \approx 1$ for contacts with conventional ferromagnets like cobalt, iron, and nickel.
This strong $Z$ dependence of the polarisation does not match the results of the Tedrow-Meservey tunneling experiments \cite{Tedrow1973} carried out in the opposite $Z \gg 1$ limit, indicating that the interface transparency affects the measured polarization \cite{Kant2002,Woods2004}.

Although the BTK $Z$ is assumed to be well understood, it is very often treated as a simple fit parameter without further consideration.
We show here for contacts between aluminium, one of the most simple superconductors, and various non-magnetic normal metals, that the $Z$  parameter is very likely not related to Fermi surface mismatch as it is understood today.

\section{Experimental}

Our point contacts were made using the shear (crossed wire) method by gently touching with one sample wire another one.
The wires had diameters of either $0.25\,$mm (all normal metals) or $0.5\,$mm (Al). 
The contacts were measured at temperatures down to 0.1\,K in the vacuum region of a dilution refrigerator.
A DC current $I$ with a small superposed AC component $dI$ is injected into 
the contact and the voltage drop $V+dV$ across the contact measured to obtain the $I(V)$ characteristics as well as the differential resistance spectrum $dV/dI(V)$.
We estimate the contact radius $a$ by the ballistic Sharvin resistance $R = 2 R_K / (ak_F)^2$ where $R_K = h/e^2$.
In free-electron approximation these metals have typically Fermi wave numbers of $k_F \approx 14\,$nm$^{-1}$ \cite{Ashcroft1976}. 
Thus a $1\,\Omega$ contact has a radius of $a \approx 16\,$nm, assuming circular symmetry, or $\sim \! 830\,{\text{nm}}^2$ cross-sectional area.
This agrees reasonably well with the $\pi a^2 R \approx 1.1\,$f$\Omega$m$^2$ CPP resistance of Al - Cu interfaces \cite{Sharma2009}.

\begin{figure}
\includegraphics[width=8.5cm]{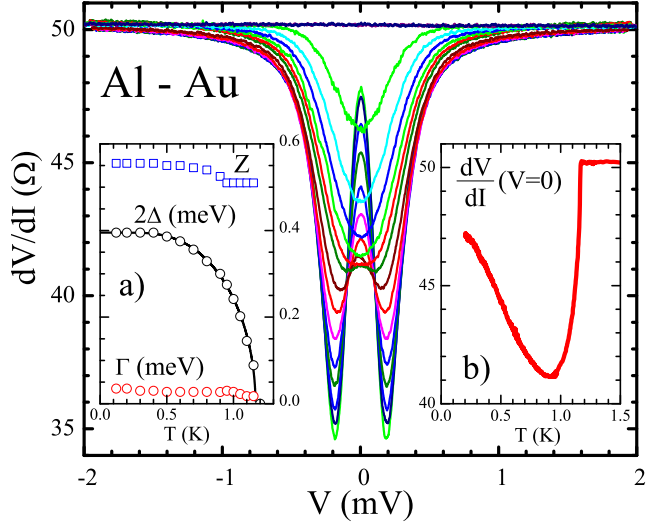}
\caption{(Color online) Differential resistance spectra $dV/dI$ versus bias voltage $V$ of an Al - Au contact in the temperature range from 0.1\,K to 0.9\,K in 0.1\,K steps and above $0.9\,$K in 0.05\,K steps. The insets show a) the superconducting energy gap $2\Delta$ together with the theoretical BCS temperature dependence (solid line), the $Z$ and the $\Gamma$ parameter as function of temperature as well as b) the temperature dependence of the zero bias resistance $dV/dI(V=0)$.}
  \label{spectra} 
  \end{figure}

\begin{figure}
\includegraphics[width=8.5cm]{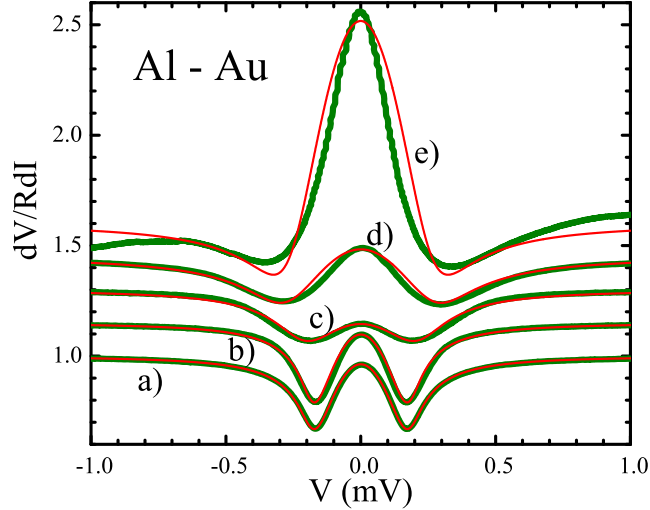}
\caption{(Color online) Typical differential resistance spectra $dV/dI$ normalized with respect to the normal contact resistance $R$ versus bias voltage $V$ of Al - Au contacts at $T = 0.1\,$K. 
The fits (thin lines) of the low-resistance contacts a) to c) match perfectly the measured curves (thick solid curves). The spectra b) to e) have been shifted vertically in units of 0.15. 
They have the following parameters: 
a) $R = 3.2\,\Omega$, $2\Delta_0 = 365\,\mu$eV, $\Gamma = 28\,\mu$eV, and $Z = 0.56$. 
b) $22.5\,\Omega$, $365\,\mu$eV, $22\,\mu$eV, and $0.55$. 
c) $482\,\Omega$, $455\,\mu$eV, $78\,\mu$eV, and $0.44$. 
d) $2.1\,$k$\Omega$, $590\,\mu$eV, $102\,\mu$eV, and $0.65$. 
e) $1.6\,$k$\Omega$, $560\,\mu$eV, $82\,\mu$eV, and $1.41$.} 
 \label{spectral-shape} 
  \end{figure}

We have chosen only contacts that had spectra with the characteristic double-minimum structure of Andreev reflection like the ones in Figures \ref{spectra} and \ref{spectral-shape}, roughly half of all contacts.
The spectra were analysed using the modified BTK theory that includes Dynes' lifetime parameter $\Gamma$ \cite{Plecenik1994}.
The normal resistance agreed with the asymptotic differential resistance at large bias voltages.
Side peaks at finite bias voltage, for example due to the self-magnetic field, were easy to recognize and did therefore not affect the analysis with respect to normal reflection. 

Figure \ref{spectra} shows the spectra of one of the Al - Au contacts as function of temperature. 
The extracted energy gap $2\Delta(T)$ follows closely the BCS temperature dependence, while $\Gamma$ and $Z$ remain nearly constant.
As long as the double-minimum structure exists, for this contact up to 0.9\,K, all three parameters can be determined independently (the error bars are smaller than the symbol sizes). 
The double minimum collapses into a single one at $T > 0.9\,$K. For those spectra we kept $Z$ constant and fitted only $2\Delta_0$ and $\Gamma$.
The contacts discussed below were measured at $T = 0.1\,$K$ \ll T_c$ to reduce thermal smearing in order to reliably determine $\Gamma$.
Figure \ref{spectral-shape} shows a selection of typical spectra over the accessible resistance range together with fit curves and the extracted parameters.
Noticeable deviations from the BTK-type fit appear only at very large resistances, that is very small contacts.

\section{Results and discussion}

\begin{figure}[t]
\includegraphics[width=8.5cm]{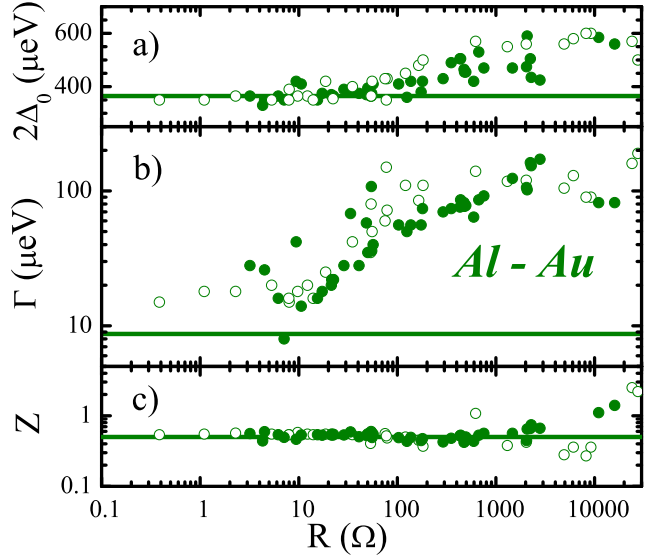}
\caption{(Color online) Properties of Al-Au contacts extracted from the modified BTK theory versus normal contact resistance $R$ at $T = 0.1\,$K: a) Superconducting energy gap $2\Delta_0$, b) Dynes lifetime parameter $\Gamma$, and c) $Z$ parameter of normal reflection. The solid lines are the BCS energy gap of Al, the thermal energy $k_B T$ at the measuring temperature, and $Z=0.5$, respectively. Different symbols mark different measurement series.}
  \label{AlOverview} 
  \end{figure}

Figure \ref{AlOverview} shows the derived parameters $2\Delta_0 = 2\Delta(T=0)$, $\Gamma$, and $Z$ as function of normal resistance $R$ for contacts between superconducting Al and Au. 
The energy gap $2\Delta_0$ agrees well with the literature value of $365\,\mu$eV from below $1\,\Omega$ up to several $10\,\Omega$. 
Then the gap increases to about twice its low-resistance value. 
The lifetime parameter $\Gamma$ is barely resolvable at contacts with small resistance because of the $\sim \! 10\,\mu$eV thermal smearing.
Between $10\,\Omega \lesssim R \lesssim 100\,\Omega$ the lifetime parameter increases almost linearly, and saturates at round $200\,\mu$eV at high resistances.

Most astonishingly, the $Z$ parameter stays constant at $\sim \! 0.5$ from below $1\,\Omega$ up to several $1000\,\Omega$, that means from contact areas of more than $1000\,$nm$^2$ to less than $1\,$nm$^2$. 
Reflection at a dielectric barrier should result in a strong variation of $Z$, depending on how a specific contact is made, because the transmission coefficient depends exponentially on the barrier width and height \cite{Simmons1964}.
The extremely small variation of $Z$ thus indicates a negligibly weak dielectric tunneling barrier.
This appears plausible for shear contacts because the two electrode wires slide along each other, scratching the surfaces and thereby removing a possible oxide or water/ice layer before the contact is formed.
$Z$ increases only in the k$\Omega$ range towards the transition to vacuum tunneling.
At such high resistances the contact consists of only a few conduction channels, each with its own reflection coefficient, where the ones with the large $Z$ are not averaged away. 
This might also contribute to the deviations from the BTK-type fits in Figure \ref{spectral-shape}.

\begin{figure}[t]
\includegraphics[width=8.5cm]{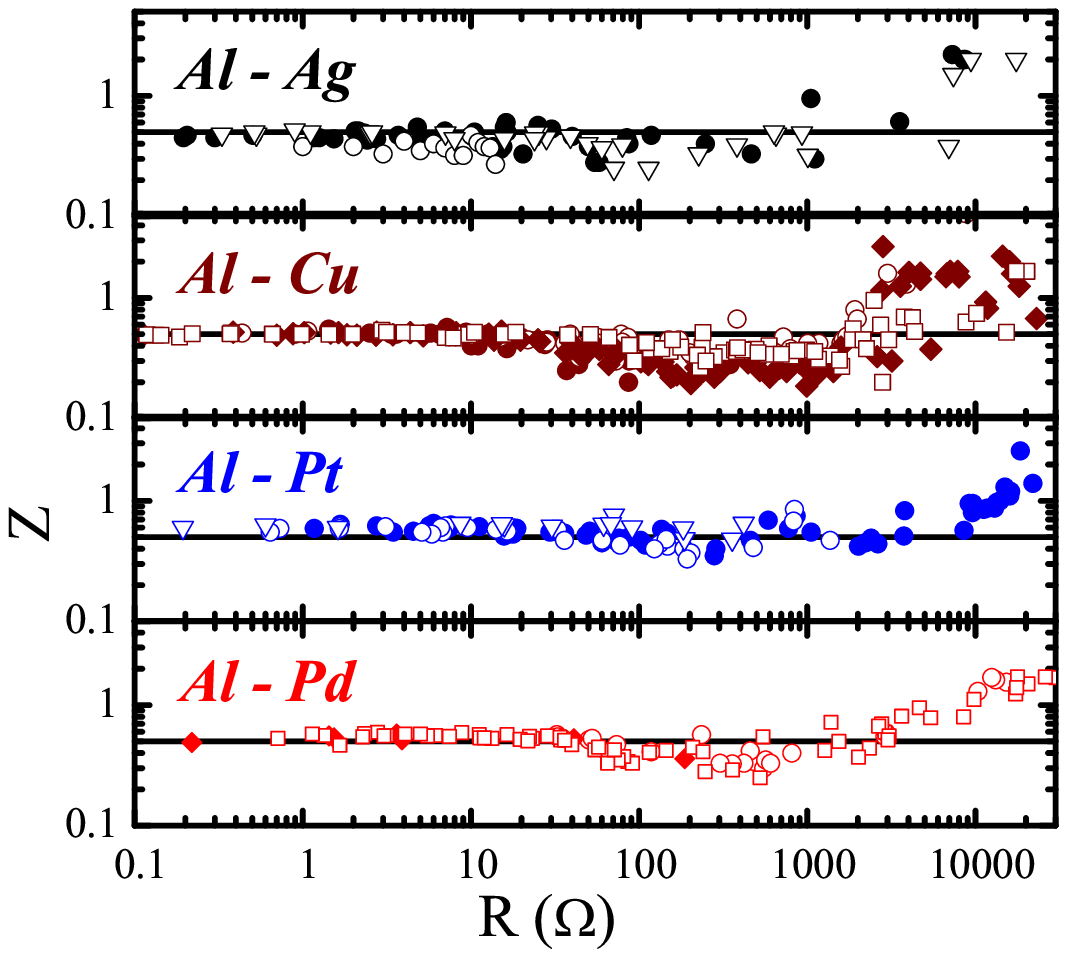}
\caption{(Color online) BTK $Z$ parameter of normal reflection versus normal resistance $R$ of Al in contact with the indicated  metals at $T = 0.1\,$K. Solid lines are $Z=0.5$ as guide to the eye. Different symbols mark different measurement series.}
  \label{Z-Parameters} 
  \end{figure}

Contacts with the other normal metals Ag, Cu, Pd, and Pt behave similarly as summarized for the $Z$ parameter in Figure \ref{Z-Parameters}. 
However, there appears to be a {\em softening} of $Z$ especially for the Al - Cu contacts at around $R \approx 100\,\Omega$, coinciding with the increasing $2\Delta_0$ and the saturation of $\Gamma$. 
The suspicion that this partial reduction of $Z$ results from electrical noise can be discarded because $Z$ recovers at higher resistances which should be even more susceptible to noise.
An alternative explanation for this behaviour might be a real size-dependent property of the contacts, analogue to the Kondo-like zero-bias anomalies \cite{Gloos2009}.

As discussed for the Al - Au contacts, the rather constant $Z$ makes a noticeable dielectric barrier unlikely also for contacts with the other normal metals. 
This leaves Fermi surface mismatch to explain the finite $Z$. 
The metal combinations presented here should have either $Z \approx 0$ in free-electron approximation or $Z \gtrsim 1$ based on proximity-effect estimates mentioned above.
Our experimental data fit neither of those two limits.
In addition, the argument that excludes tunneling also holds for Fermi surface mismatch because we can not control how the crystallites that actually form the contact are oriented.
Can it be that Fermi surface mismatch does not contribute to normal reflection as measured by Andreev reflection?

Electrons as well as Andreev-reflected holes cross a real dielectric tunneling barrier with a certain probability, while the remainder is normally reflected.
This allows to extract the transmission or the reflection coefficient from a single Andreev-reflection spectrum.
Normal reflection due to Fermi surface mismatch works differently: a certain kind of the incident electrons - those with the wrong wave number and direction of incidence - can not cross the interface.
Since only electrons that have been transmitted can be Andreev-reflected, the retro-reflected holes have already the right properties to be transmitted back through the interface.
If this is correct then Fermi surface mismatch can not be resolved using Andreev-reflection spectroscopy.

\section{Conclusion}
We have found that the BTK $Z$ parameter of interfaces with superconducting Al does neither depend sensitively on the size of the contacts nor on the normal metal.
Its tiny variation over a vast range of contact areas has lead us to conclude the absence of a dielectric tunneling barrier. 
The same argument holds for Fermi surface mismatch because of the uncontrollable orientation of the crystallites that form the contact. 
Moreover, the average experimental $Z \approx 0.5$ disagrees with the free-electron prediction ($Z \approx 0$) and with the results of the proximity studies ($Z \ge 1$).
The experiments on reflection at or transmission through normal-superconducting interfaces and their interpretation remind us that different methods can yield different results for the same physical property. 
In the end we are left with a new -- but at the same time old -- mystery of the origin of the Andreev reflection double minimum anomaly described by $Z$.

\begin{acknowledgments}
E. T.  acknowledges a two-year grant from the Graduate School of Materials Research (GSMR), 20014 Turku, Finland. We thank the Jenny and Antti Wihuri Foundation for financial support.
\end{acknowledgments}

\end{document}